\documentclass[12pt]{article}
\usepackage{amsmath,amssymb}
\usepackage{graphicx}
\usepackage{cite,fancyhdr}
\usepackage[affil-it]{authblk}
\usepackage[left=2cm,right=2cm,top=2cm,bottom=2cm,a4paper]{geometry}

\begin{document}

%\allowdisplaybreaks
\setcounter{footnote}{0}
\setcounter{figure}{0}
\setcounter{table}{0}

\title{\bf \large 
Focus Point Gauge Mediation without a Severe Fine-tuning
 }
\author[1*]{{\normalsize Tsutomu T. Yanagida}}
\author[2]{{\normalsize Norimi Yokozaki}}
\affil[1]{\small 
Kavli IPMU (WPI), UTIAS, University of Tokyo, 
Kashiwa  277-8583, Japan}
\affil[2]{\small 
Department of Physics, Tohoku University,  
Sendai, Miyagi 980-8578, Japan}

\date{}

\maketitle

\thispagestyle{fancy}
\rhead{IPMU18-0144, TU-1072}
\cfoot{\thepage}
\renewcommand{\headrulewidth}{0pt}

\begin{abstract}
\noindent
We consider focus point gauge mediation within the framework of the next-to-minimal supersymmetric standard model, which substantially reduces the degree of fine-tuning for the electroweak symmetry breaking.
The milder fine-tuning is realized by a messenger field in the adjoint representation of $SU(5)$ gauge group with $SU(3)_c$ octet being heavy. Our model has a simple ultraviolet completion. 
The fine-tuning measure $\Delta$ can be as small as 40-50 without any contradiction with LHC constraints.

\end{abstract}

\renewcommand{\thefootnote}{\fnsymbol{footnote}}
\footnote[0]{*Hamamatsu Professor}
\renewcommand{\thefootnote}{\arabic{footnote}}

\newpage

\section{Introduction}
Supersymmetric (SUSY) extensions of the standard model (SM) are the most attractive models beyond the SM, 
since they not only explain the observed mass of the Higgs boson~\cite{Aad:2015zhl} naturally, but also provide us with a consistent framework of the unification of all the known gauge coupling constants at the scale around $10^{16}$\,GeV, called the grand unified theory (GUT) scale. 
However, there is a serious problem in the SUSY SM, that is, we have too large flavor changing neutral currents (FCNCs)~\cite{Gabbiani:1996hi}. 
There have been observed, so far, only two solutions to this problem with generic K{\"a}hler potential.
One is gauge mediation~\cite{Dine:1993yw,Dine:1994vc,Dine:1995ag
} and the other high scale SUSY with the gravitino mass $\gtrsim$\,100-1000\,TeV~\cite{ArkaniHamed:2004fb,Giudice:2004tc,ArkaniHamed:2004yi,Wells:2004di,pure1,Giudice:2011cg,Hall:2011jd,pure2,pure3}. Furthermore, the former does not have a serious cosmological problem so called ``Polonyi Problem"\cite{polonyi1}.\footnote{Pure gravity mediation~\cite{pure1,pure2,pure3} which belongs to high scale SUSY does not have the Polonyi field and hence it is free from the cosmological problems.}

The purpose of this paper is to discuss the fine-tuning problem for the electroweak symmetry breaking (EWSB) scale 
in gauge mediation models. 
We first show that minimal gauge mediation in the minimal SUSY SM (MSSM)
already needs a severe fine-tuning as $\Delta \gtrsim1500$ at the present, 
where $\Delta$ shows the sensitivity of the $Z$-boson mass scale to fundamental mass parameters~\cite{ft_measure1,ft_measure2}.
This is because we need large stop masses to explain the observed Higgs boson mass of 125 GeV~\cite{Okada:1990vk,Ellis:1990nz,Haber:1990aw,Okada:1990gg,Ellis:1991zd}. 
Therefore, we introduce a singlet chiral multiplet to the MSSM (NMSSM) to lower the stop masses while keeping the Higgs boson mass. However, we find that a fine-tuning of $\Delta \gtrsim 300$ is still required.
Finally, we invoke focus point gauge mediation proposed by Fukuda et al. \cite{Fukuda:2015pra} (see also \cite{Agashe:1999ct,Brummer:2012zc,Brummer:2013yya} for earlier attempts) some time ago and show that we do not need such a severe fine-tuning.\footnote{With a specific form of the K{\"a}her potential, such as a sequestered K{\"a}hler potential, the too large FCNCs are also avoided. In this case, focus point gaugino mediation~\cite{Yanagida:2013ah,Yanagida:2013uka,Yanagida:2014cxa} is considered, which ameliorates the fine-tuning of the EWSB scale. The reduction of fine-tuning is a consequence of non-universal gaugino masses at the GUT scale (see for example Refs.~\cite{Kane:1998im,BasteroGil:1999gu,Abe:2007kf,Martin:2007gf,Baer:2007xd,Horton:2009ed,Younkin:2012ui,Antusch:2012gv,Gogoladze:2012yf,Spies:2013fba,Martin:2013aha}).} In fact, we find a wide parameter region with $\Delta \simeq 40$-50 in our model. 
We discuss predictions and testability of the model in conclusion.

%#######################
\section{Minimal gauge mediation in MSSM and the fine-tuning problem}

In this section, we show that the minimal gauge mediation model in the MSSM requires a very high degree of fine-tuning due to large masses for colored SUSY particles.
The superpotential for ${\bf 5}$ and $\bar{\bf 5}$ messengers is given by
\begin{eqnarray}
W = \lambda_L^I Z \Psi_{L}^I \Psi_{\bar{L}}^I + \lambda_D^I Z \Psi_{D}^I \Psi_{\bar{D}}^I,
\end{eqnarray}
where $I=1\dots N_5$, $\Psi_L^I$ and $\Psi_D^I$ are $SU(2)_L$ doublet and $SU(3)_c$ triplet messengers, respectively, and  
$Z$ is a SUSY breaking field which has non-vanishing vacuum expectation values (VEVs): $Z=M + F_Z \theta^2$. 
Since $\Psi_L^I$ and $\Psi_{\bar D}^I$ consists of a complete multiplet of $SU(5)$ GUT gauge group, $\bar{\bf 5}$,
$\lambda_L^I = \lambda_D^I$ at the GUT scale is expected. 
By assuming all the couplings are of the same order, we define the messenger scale as $M_{\rm mess}=\lambda_5 M$, where $\lambda_5$ is a coupling $\lambda_{L}^I$ or $\lambda_D^I$.

%============
\begin{figure}[t]
 \begin{center}
   \includegraphics[width=80mm]{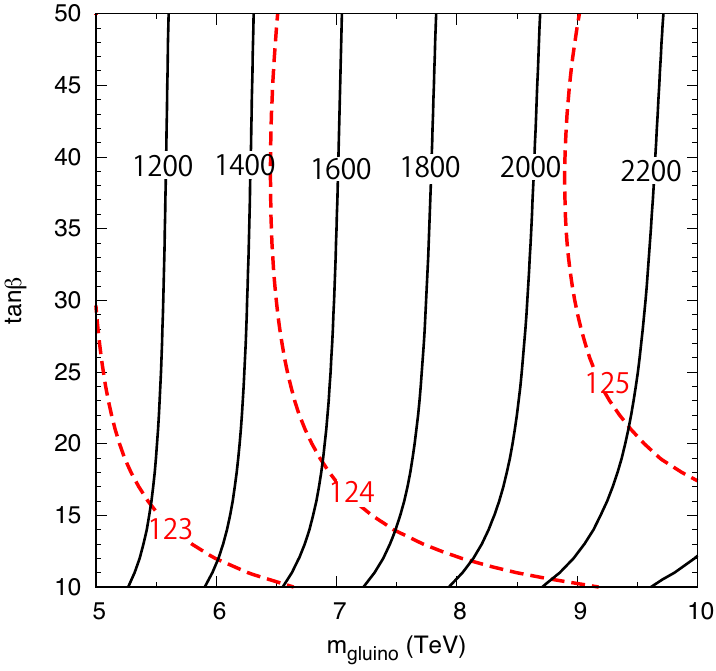}
   \includegraphics[width=80mm]{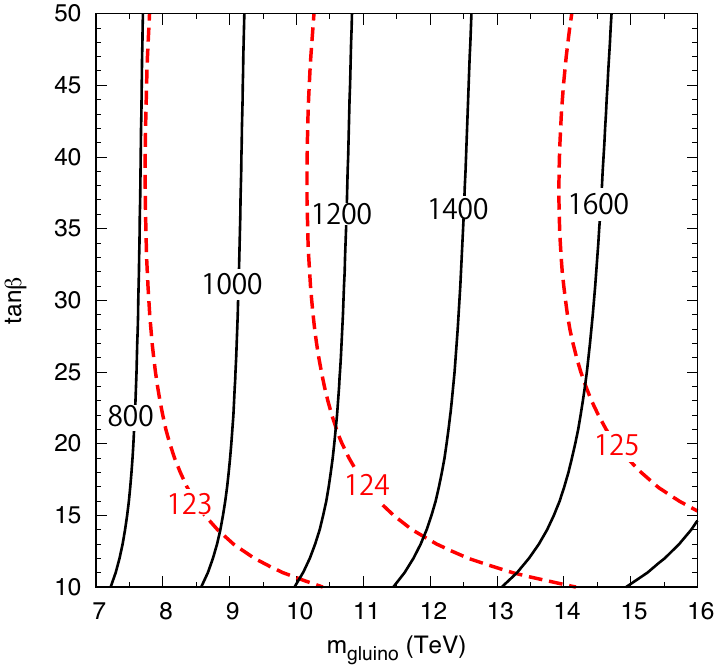}
 \end{center}
  \caption{The contours of $\Delta$ (black solid) and $m_h$ (red dashed), where $m_h$ is shown in units of GeV. 
  We take $M_{\rm mess}=1500$\,TeV and $N_5=1$ ($M_{\rm mess}=700$\,TeV and $N_5=4$) in the left (right) panel. 
Here, $\alpha_s (m_Z) =0.1181$ and $m_t({\rm pole})=173.34$\,GeV. }
 \label{fig:del_mssm}
\end{figure}
%============

After integrating out the messenger fields, we obtain the soft SUSY breaking masses for the MSSM particles as
\begin{eqnarray}
M_{\tilde b} = \frac{g_1^2}{16\pi^2} (N_5 \Lambda),\,
M_{\tilde w} =\frac{g_2^2}{16\pi^2} (N_5\Lambda), \,
M_{\tilde g} = \frac{g_3^2}{16\pi^2} (N_5\Lambda),
\end{eqnarray}
and
\begin{eqnarray}
m_{\tilde Q}^2 &\simeq& \frac{N_5}{256\pi^4} \left[ \frac{8}{3} g_3^4  + \frac{3}{2} g_2^4  + \frac{6}{5}g_1^4 \left(\frac{1}{6^2}\right) \right] \Lambda^2 , \nonumber \\
m_{\tilde U}^2 &\simeq& \frac{N_5}{256\pi^4} \left[ \frac{8}{3} g_3^4 + \frac{6}{5}g_1^4 \left(\frac{2}{3}\right)^2 \right] \Lambda^2 , \nonumber \\
m_{\tilde D}^2 &\simeq& \frac{N_5}{256\pi^4} \left[ \frac{8}{3} g_3^4  + \frac{6}{5}g_1^4 \left(\frac{1}{3^2}\right) \right] \Lambda^2 , \nonumber \\
m_{\tilde L}^2 &\simeq& \frac{N_5}{256\pi^4} \left[ \frac{3}{2} g_2^4  + \frac{6}{5}g_1^4  \left(\frac{1}{2^2}\right) \right] \Lambda^2 , \nonumber \\
m_{\tilde E}^2 &\simeq& \frac{N_5}{256\pi^4} \left[ \frac{6}{5}g_1^4  \right] \Lambda^2 , \nonumber \\
m_{H_u}^2 &=& m_{H_d}^2 = m_{\tilde L}^2,
\end{eqnarray}
where $M_{\tilde b}$, $M_{\tilde w}$ and $M_{\tilde g}$ are the bino, wino and gluino masses, respectively; $m_{\tilde Q}$, $m_{\tilde U}$ and $m_{\tilde D}$ are squark masses; $m_{\tilde L}$ and $m_{\tilde E}$ are slepton masses; $m_{H_u}^2$ and $m_{H_d}^2$ are soft masses for the up-type and down-type Higgs, respectively; $g_1$, $g_2$ and $g_3$ are gauge coupling constants of $U(1)_Y$, $SU(2)_L$ and $SU(3)_c$. Here, $\Lambda = F_Z/M$, which controls the overall mass scale.

The fine-tuning is estimated using the following measure~\cite{ft_measure1,ft_measure2}:
\begin{eqnarray}
\Delta = {\rm max} \left|\frac{\partial \ln m_Z}{\partial \ln a_i}\right|, \,
a_i = \left\{ |\mu|, M, |F_Z|, |B_\mu (M_{\rm mess})| \right\},
\end{eqnarray}
where $\mu$ is a higgsino mass and $B_\mu(M_{\rm mess})$ is a Higgs $B$-term at the messenger scale, which is defined as $V \ni B_\mu H_u H_d + h.c.$ We consider $a_i$ is a fundamental mass parameter, and check the sensitivity of the $Z$-boson mass $m_Z$ to $a_i$.

In Fig.~\ref{fig:del_mssm}, we show contours of the Higgs boson mass $m_h$ and $\Delta$ on $m_{\rm gluino}$-$\tan\beta$ plane, where $m_h$ is shown in units of GeV. Here, $m_{\rm gluino}$ is a physical gluino mass and $\tan\beta$ 
is a ratio of the Higgs VEVs, $\left<H_u^0\right>/\left<H_d^0\right>$.
SUSY mass spectra and $m_h$ are calculated using {\tt SPheno-4.0.3}~\cite{Porod:2003um,Porod:2011nf}. 
In the left (right) panel, we take $M_{\rm mess}=1500$\,TeV and $N_5=1$ ($M_{\rm mess}=700$\,TeV and $N_5=4$). In the case of $N_5=4$, the fine-tuning is slightly milder than the case of $N_5=1$ for fixed $m_h$, as the messenger scale can be lower. However, even in this case, the fine-tuning of $\Delta \gtrsim 1500$ is required to explain the Higgs boson mass of 125\,GeV. The stop mass scale, $\sqrt{m_{\tilde{Q}_3} m_{\tilde{U}_3}}$, for $m_h=125$\,GeV is around 10\,TeV in both cases. Note that $\Delta$ is dominated by the sensitivity to $\mu$-parameter.

\section{Minimal gauge mediation in NMSSM}

In the MSSM, the very high degree of fine-tuning is required since the stops need to be heavy as $\sim 10$\,TeV to explain the observed Higgs boson mass. Thus, we introduce a singlet chiral multiplet $S$ to the MSSM, lowering the stop masses. In this section, only one pair of ${\bf 5}$ and $\bar{\bf 5}$ messengers is introduced, i.e. $N_5=1$. This is because a stau becomes the next-to-lightest SUSY particle for larger $N_5$,  which is quite severely constrained by LHC experiments~\cite{stau_lhc}. Here, we assume the stau is long-lived with the gravitino heavier than $\mathcal{O}$(10)\,keV.\footnote{The gravitino lighter than about 10\,keV is strongly constrained by the Lyman-$\alpha$ forest data~\cite{Baur:2015jsy}, 
if it is a dominant component of the dark matter.}

The relevant superpotential and scalar potential in the NMSSM are written as
\begin{eqnarray}
W = \lambda S H_u H_d + \xi_F S + \frac{1}{2} \mu' S^2,
\end{eqnarray}
and
\begin{eqnarray}
V_{\rm soft} = m_S^2 |S|^2 + (A_\lambda \lambda S H_u H_d + \xi_S S+ \frac{1}{2}m_S'^2 S^2 +  h.c.),
\end{eqnarray}
respectively. The mass parameters in the superpotential are $|\xi_F|^{1/2} \sim \mu' \sim \mathcal{O}(1)$\,TeV, and we assume $S^3$ term is suppressed by a symmetry. In order to realize the correct EWSB, non-zero soft SUSY breaking mass parameters in the scalar potential are required. In our setup, the correct EWSB is realized by the tadpole, $\xi_S \sim \mathcal{O}({\rm TeV}^3)$. 
We take $m_S^2 = 10^6$\,GeV$^2$ and $A_\lambda=m_S'^2=0$ at the messenger scale.
An explicit example model inducing these soft SUSY breaking parameters is shown in Appendix A.

In this setup, the fine-tuning is estimated using
\begin{eqnarray}
\Delta = {\rm max} \left|\frac{\partial \ln m_Z}{\partial \ln a'_i}\right|, \,
a'_i = \left\{ |\mu'|, |\xi_F|, |\xi_S(M_{\rm mess})|, |m_S^2(M_{\rm mess})|, M, |F_Z|| \right\}, \label{eq:del_nmssm}
\end{eqnarray}
where $a'_i$ is considered to be a fundamental mass parameter in this setup.

%============
\begin{figure}[t]
 \begin{center}
   \includegraphics[width=80mm]{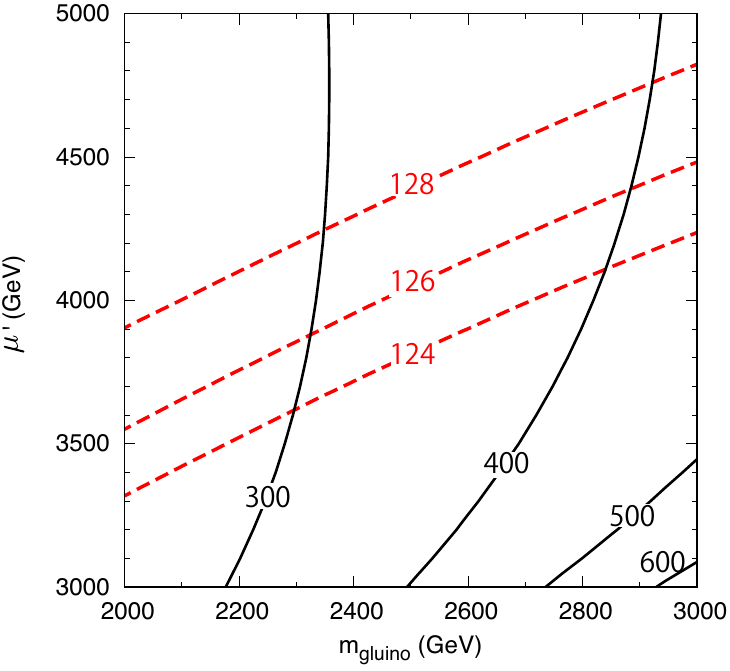}
 \end{center}
  \caption{The contours of $\Delta$ (black solid) and $m_h$ (red dashed), where $m_h$ is shown in units of GeV. 
  We take $M_{\rm mess}=500$\,TeV, $\lambda=0.9$ and $\tan\beta=4.0$. The other parameters are same as in Fig.~\ref{fig:del_mssm}.}
 \label{fig:del_nmssm}
\end{figure}
%============

We show contours of $\Delta$ and $m_h$ in Fig.~\ref{fig:del_nmssm} on $m_{\rm gluino}$-$\mu'$ plane. 
SUSY mass spectra and $m_h$ are estimated using {\tt NMSSMTools 5.2.0}~\cite{Ellwanger:2008py,Allanach:2015mwa
}. 
The coupling $\lambda$ is defined at the stop mass scale while $\mu'$ is given at the messenger scale.
In the framework of the NMSSM, the Higgs boson mass of 125\,GeV is easily explained thanks to the $F$-term contribution from $\lambda S H_u H_d$ 
with $\mu'$ of $\mathcal{O}(1)$\,TeV; therefore, the minimum of $\Delta$ is not determined by $m_h$ but current LHC constrains on gluino and squark masses~\cite{Aaboud:2017vwy}. By considering these constraints, $\Delta$ around 300 is required in this framework.

\section{Focus point gauge mediation in NMSSM}

The fine-tuning of the EWSB scale is further relaxed when we adapt focus point gauge mediation~\cite{Fukuda:2015pra}, 
which is realized in $SU(5) \times U(3)_H$ product group unification (PGU)~\cite{Yanagida:1994vq,Izawa:1997he}.
In $SU(5) \times U(3)_H$ PGU, the doublet-triplet splitting problem in GUTs is elegantly solved.

To realize focus point gauge mediation, a messenger superfield in the adjoint ${\bf 24}$ representation of $SU(5)$ is introduced. The messenger field of $SU(3)_c$ octet in the ${\bf 24}$ representation becomes much heavier than others, due to a Dirac mass with another $SU(3)_c$ octet chiral field, which belongs to the adjoint representation of $SU(3)_H ( \subset U(3)_H)$ before $SU(5) \times U(3)_H$ is broken to the SM gauge group. 
Then, the low-energy Lagrangian in the messenger sector is given by
\begin{eqnarray}
W = \lambda_X Z X \bar X + \lambda_3 Z {\rm Tr}(\Sigma_3^2),
\end{eqnarray}
where $X$, $\bar X$ and $\Sigma_3$ correspond to $({\bf 3},{\bf 2})$, $(\bar{\bf 3},{\bf 2})$ and $({\bf 1},{\bf 3})$ of the $SU(3)_c \times SU(2)_L$ gauge group, and $U(1)_Y$ charges of $X$ and $\bar X$ are -5/6 and 5/6, respectively. Note that the gauge coupling unification is still maintained with contributions from the $SU(3)_c$ octet fields~\cite{Fukuda:2015pra}.

After integrating out the messenger fields, we obtain
\begin{eqnarray}
M_{\tilde b} = \frac{g_1^2}{16\pi^2} (5 \Lambda), \,
M_{\tilde w} = \frac{g_2^2}{16\pi^2} (5 \Lambda) , \,
M_{\tilde g} = \frac{g_3^2}{16\pi^2} (2 \Lambda) ,
\end{eqnarray}
and 
\begin{eqnarray}
m_{\tilde Q}^2 &\simeq& \frac{1}{256\pi^4} \left[ \frac{8}{3} g_3^4 (2 \Lambda^2) + \frac{3}{2} g_2^4 (5\Lambda^2) + \frac{6}{5}g_1^4 (5\Lambda^2) \frac{1}{6^2} \right] \nonumber \\
m_{\tilde U}^2 &\simeq& \frac{1}{256\pi^4} \left[ \frac{8}{3} g_3^4 (2 \Lambda^2) + \frac{6}{5}g_1^4 (5\Lambda^2) \left(\frac{2}{3}\right)^2 \right] \nonumber \\
m_{\tilde D}^2 &\simeq& \frac{1}{256\pi^4} \left[ \frac{8}{3} g_3^4 (2 \Lambda^2)  + \frac{6}{5}g_1^4 (5\Lambda^2) \frac{1}{3^2} \right] \nonumber \\
m_{\tilde L}^2 &\simeq& \frac{1}{256\pi^4} \left[ \frac{3}{2} g_2^4 (5\Lambda^2) + \frac{6}{5}g_1^4 (5\Lambda^2) \frac{1}{2^2} \right] \nonumber \\
m_{\tilde E}^2 &\simeq& \frac{1}{256\pi^4} \left[ \frac{6}{5}g_1^4 (5\Lambda^2) \right] \nonumber \\
m_{H_u}^2 &=& m_{H_d}^2 = m_{\tilde L}^2.
\end{eqnarray}
Notice that $SU(2)_L$ contributions to the gaugino and sfermion masses correspond to the effective messenger number  of five, while $SU(3)_c$ contributions correspond to the effective messenger number of two. These larger $SU(2)_L$ contributions allow the small fine-tuning. Since the focus point leading to the small fine-tuning is fixed by the representations of $SU(5)$ and $U(3)_H$, the scenario is highly predictive and robust.
As in the previous section, we set $m_S^2=10^6$\,GeV$^2$ and $A_\lambda=m_S'=0$.

%============
\begin{figure}[t]
 \begin{center}
   \includegraphics[width=80mm]{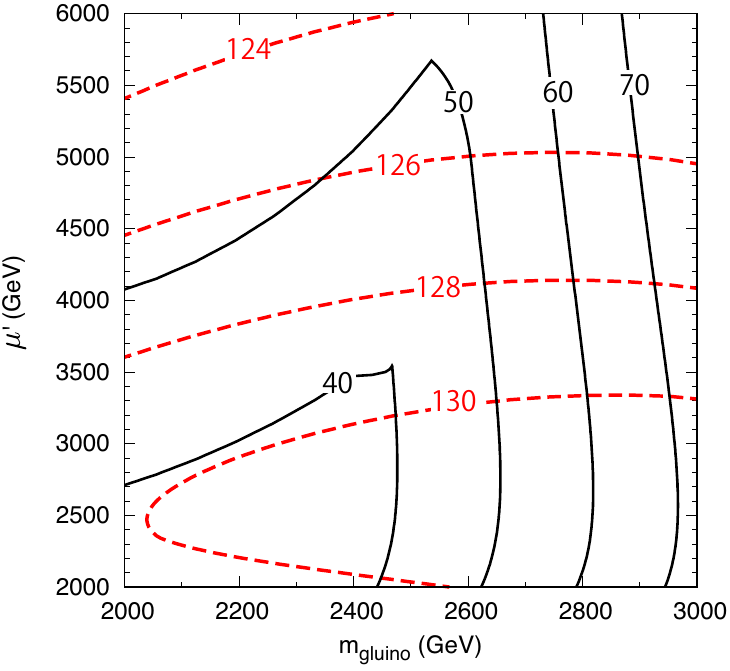}
 \end{center}
  \caption{The contours of $\Delta$ (black solid) and $m_h$ (red dashed), where $m_h$ is shown in units of GeV. 
  The parameters are same as in Fig.~\ref{fig:del_nmssm}.}
 \label{fig:del_focus}
\end{figure}
%============

\begin{table*}[!t]
\caption{Mass spectra in sample points. Here, $\mu_{\rm eff}= \lambda \left<S\right>$.
}
\label{tab:sample}
%\begin{table}[]
\begin{center}
\begin{tabular}{|c||c|c|c|c|}
\hline
Parameters & Point {\bf I} & Point {\bf II} & Point {\bf III}   \\
\hline
$M_{\rm mess} $\,(TeV) & 1000  & 500  & $1000$\\
$\Lambda $\,(TeV) & 183  & 200   & 200\\
$\lambda$  & $0.8$  & $0.9$  & 0.8 \\
$\mu'(M_{\rm mess})$\,(GeV)  & $2600$  & $4600$ & 2200 \\
$\tan\beta$ & 4 & 4 &4\\
\hline
%\hline
%
Particles & Mass (GeV) & Mass (GeV)  & Mass (GeV)\\
\hline
$\tilde{g}$ & 2600 & 2880 & 2830\\
$\tilde{q}$ & 2700-3080 & 2960-3360 & 2930-3340\\
$\tilde{t}_{1,2}$ & 2420,\,2950 & 2690,\,3230  & 2630,\,3200\\
$\tilde{\chi}_{1}^\pm$ & 597 & 486   & 596\\
$\tilde{\chi}_{2}^\pm$ & 2360 & 2640 & 2580\\
$\tilde \chi_1^0$         & 591 & 482   & 590\\
$\tilde{\chi}_2^0$        & 605 & 492  & 604\\
$\tilde \chi_3^0$         & 1280 & 1430 & 1400\\
$\tilde{\chi}_4^0$       & 2340 & 2640   & 1990\\
$\tilde{\chi}_5^0$       & 2360 & 4090   & 2580\\
$\tilde{e}_{L, R}$       & 1560,\,760 & 1680,\,815  & 1700,\,829\\
%$\tilde{\tau}_{1,2}$     & 733,1520&  535,\,935  & 501,\,1030 \\
$H^{\pm}$ & 1630 & 1680 & 1760 \\
$h_{\rm SM\mathchar`-like}$ & 127.2 &  126.9  & 126.4 \\
\hline
$\mu_{\rm eff}$\,(GeV)  & 586  & 475  & 584\\
$\Delta$  & 44  & 68  & 57\\
\hline
\end{tabular} \label{table:mass}
%\end{table}
\end{center}
\end{table*}

In Fig.~\ref{fig:del_focus}, we show contours of $\Delta$ (cf.\,Eq.(\ref{eq:del_nmssm})) and $m_h$ in focus point gauge mediation within the framework of the NMSSM. As in the previous case, SUSY mass spectra and $m_h$ are computed using {\tt NMSSMTools} 
with modifications to incorporate focus point gauge mediation.
One can see that $\Delta$ is reduced to 40-50 in this framework for the gluino mass of 2.4-2.6\,TeV. 
%In this setup, $\Delta$ is dominated by the sensitivity to $M$.

Finally, some examples of mass spectra are shown in Table.~\ref{table:mass}. 
For all the benchmark points, 
the higgsino is the next-lightest SUSY particle, which is assumed to be stable in the collider time scale.
%The fine-tuning is small as $\Delta < 50$. 
Due to the small $\tan\beta$, smuon and stau masses are almost degenerate with selectron masses.

\section{Conclusions}\label{sec:discuss}

We have considered focus point gauge mediation in the NMSSM and shown that the fine-tuning $\Delta$ is significantly reduced compared to the minimal gauge mediation in the MSSM.
It has been found that there exists a wide range of parameter with $\Delta \simeq 40$-50, where the minimum value of $\Delta$ is determined by the LHC constraints on the gluino and squark masses. 
The EWSB is correctly explained 
with the SUSY breaking tadpole term in the scalar potential for the singlet Higgs. 
An explicit example model generating the tadpole is shown in Appendix A.

In our scenario, the gluino dominantly decays into the higgsino, third generation quark and antiquark with the large top-Yukawa coupling. 
In this case, the LHC constraint on the gluino mass is about 2\,TeV for decoupled squarks~\cite{Aaboud:2017hrg}.
On the other hand, the lower-limit on the squark masses is more severe: 
the limit on the degenerated squark masses in the simplified model analysis is around 2.6\,TeV for the gluino mass of 2.6\,TeV~\cite{Aaboud:2017vwy}. Although this limit is not directly applicable to our model, the region with $\Delta<50$ is expected to be tested in near future.

%\vspace{20pt}
%Memo: The gluino dominantly decays into higgsino and top/bottom pairs (3 body decays), picking up the top-Yukawa coupling.
%The left-handed squarks dominantly decay into gluino and quarks.
%The right-handed squarks dominantly decay into bino and quarks or gluino and quarks.
%The bino decays into right-handed sleptons and leptons.
%The right-handed sleptons decay into neutral higgsino and leptons.

\section*{ Acknowledgements}
We thank Chengcheng Han for useful discussions about LHC constraints.
This work is supported by JSPS KAKENHI Grant Numbers 
JP26104001 (T.T.Y), JP26104009 (T.T.Y), JP16H02176 (T.T.Y),
JP17H02878 (T.T.Y), JP15H05889 (N.Y.),
JP15K21733 (N.Y.), JP17H05396 (N.Y.), JP17H02875 (N.Y.), and by World
Premier International Research Center Initiative (WPI Initiative),
MEXT, Japan (T.T.Y.).

\appendix 

\section{Generation of tadpole}

To generate the tadpole term realizing the EWSB, we consider the following superpotential:
\begin{eqnarray}
W = \lambda_{12}' S \Psi_1 \bar\Psi_2 + \mu_{12}' \Psi_1 \bar\Psi_2 +   \lambda_1 Z\Psi_1 \bar\Psi_1 + M_{2} \Psi_2 \bar\Psi_2, 
\end{eqnarray}
where $\left<Z\right> = M + F_Z \theta^2$, $\mu_{12}' \sim 1$\,TeV, and $\Psi_{1,2}$ is assumed to be charged under a gauged or global  $SU(N)$.  We take $\lambda_1=1$ and $M_2=M$ for simplicity. After integrating out $\Psi_{1,2}$ and  $\bar \Psi_{1,2}$, we obtain
\begin{eqnarray}
\xi_S \simeq N \frac{\lambda_{12}' \mu_{12}' |F_Z|^2}{192 \pi^2 M^2},
\end{eqnarray}
and
\begin{eqnarray}
m_S^2 \simeq N \frac{\lambda_{12}'^2 |F_Z|^2}{192\pi^2 M^2} = \frac{\lambda_{12}' \xi_S}{\mu_{12}'}.
\end{eqnarray}
Other soft SUSY breaking terms can be suppressed as
\begin{eqnarray}
 m_S'^2/\mu' = 2 A_\lambda \simeq  N \frac{\lambda_{12}'^2 F_Z }{64\pi^2 M},
\end{eqnarray}
which are irrelevant to our discussion.

\providecommand{\href}[2]{#2}\begingroup\raggedright\endgroup


\begin{thebibliography}{99}

\bibitem{Aad:2015zhl} 
  G.~Aad {\it et al.} [ATLAS and CMS Collaborations],
  %``Combined Measurement of the Higgs Boson Mass in $pp$ Collisions at $\sqrt{s}=7$ and 8 TeV with the ATLAS and CMS Experiments,''
  Phys.\ Rev.\ Lett.\  {\bf 114}, 191803 (2015)
%  doi:10.1103/PhysRevLett.114.191803
  [arXiv:1503.07589 [hep-ex]].
  %%CITATION = doi:10.1103/PhysRevLett.114.191803;%%
  %1176 citations counted in INSPIRE as of 03 Sep 2018


 %\cite{Gabbiani:1996hi}
\bibitem{Gabbiani:1996hi} 
  F.~Gabbiani, E.~Gabrielli, A.~Masiero and L.~Silvestrini,
  %``A Complete analysis of FCNC and CP constraints in general SUSY extensions of the standard model,''
  Nucl.\ Phys.\ B {\bf 477}, 321 (1996)
%  doi:10.1016/0550-3213(96)00390-2
  [hep-ph/9604387].
  %%CITATION = doi:10.1016/0550-3213(96)00390-2;%%
  %1293 citations counted in INSPIRE as of 03 Sep 2018
 
 



%\cite{Dine:1993yw}
\bibitem{Dine:1993yw} 
  M.~Dine and A.~E.~Nelson,
  %``Dynamical supersymmetry breaking at low-energies,''
  Phys.\ Rev.\ D {\bf 48}, 1277 (1993)
%  doi:10.1103/PhysRevD.48.1277
  [hep-ph/9303230].
  %%CITATION = doi:10.1103/PhysRevD.48.1277;%%
  %1089 citations counted in INSPIRE as of 03 Sep 2018

\bibitem{Dine:1994vc} 
  M.~Dine, A.~E.~Nelson and Y.~Shirman,
  %``Low-energy dynamical supersymmetry breaking simplified,''
  Phys.\ Rev.\ D {\bf 51}, 1362 (1995)
%  doi:10.1103/PhysRevD.51.1362
  [hep-ph/9408384].
  %%CITATION = doi:10.1103/PhysRevD.51.1362;%%
  %1314 citations counted in INSPIRE as of 03 Sep 2018

\bibitem{Dine:1995ag} 
  M.~Dine, A.~E.~Nelson, Y.~Nir and Y.~Shirman,
  %``New tools for low-energy dynamical supersymmetry breaking,''
  Phys.\ Rev.\ D {\bf 53}, 2658 (1996)
%  doi:10.1103/PhysRevD.53.2658
  [hep-ph/9507378].
  %%CITATION = doi:10.1103/PhysRevD.53.2658;%%
  %1346 citations counted in INSPIRE as of 03 Sep 2018
  
  
  
  \bibitem{ArkaniHamed:2004fb} 
  N.~Arkani-Hamed and S.~Dimopoulos,
  %``Supersymmetric unification without low energy supersymmetry and signatures for fine-tuning at the LHC,''
  JHEP {\bf 0506}, 073 (2005)
%  doi:10.1088/1126-6708/2005/06/073
  [hep-th/0405159].
  %%CITATION = doi:10.1088/1126-6708/2005/06/073;%%
  %1055 citations counted in INSPIRE as of 03 Sep 2018
  
  
  \bibitem{Giudice:2004tc} 
  G.~F.~Giudice and A.~Romanino,
  %``Split supersymmetry,''
  Nucl.\ Phys.\ B {\bf 699}, 65 (2004)
  Erratum: [Nucl.\ Phys.\ B {\bf 706}, 487 (2005)]
%  doi:10.1016/j.nuclphysb.2004.11.048, 10.1016/j.nuclphysb.2004.08.001
  [hep-ph/0406088].
  %%CITATION = doi:10.1016/j.nuclphysb.2004.11.048, 10.1016/j.nuclphysb.2004.08.001;%%
  %823 citations counted in INSPIRE as of 03 Sep 2018

%\cite{ArkaniHamed:2004yi}
\bibitem{ArkaniHamed:2004yi} 
  N.~Arkani-Hamed, S.~Dimopoulos, G.~F.~Giudice and A.~Romanino,
  %``Aspects of split supersymmetry,''
  Nucl.\ Phys.\ B {\bf 709}, 3 (2005)
%  doi:10.1016/j.nuclphysb.2004.12.026
  [hep-ph/0409232].
  %%CITATION = doi:10.1016/j.nuclphysb.2004.12.026;%%
  %598 citations counted in INSPIRE as of 03 Sep 2018

  
  \bibitem{Wells:2004di} 
  J.~D.~Wells,
  %``PeV-scale supersymmetry,''
  Phys.\ Rev.\ D {\bf 71}, 015013 (2005)
%  doi:10.1103/PhysRevD.71.015013
  [hep-ph/0411041].
  %%CITATION = doi:10.1103/PhysRevD.71.015013;%%
  %264 citations counted in INSPIRE as of 03 Sep 2018

  
      %Pure gravity mediation
    \bibitem{pure1} 
  M.~Ibe, T.~Moroi and T.~T.~Yanagida,
  %``Possible Signals of Wino LSP at the Large Hadron Collider,''
  Phys.\ Lett.\ B {\bf 644}, 355 (2007)
  [hep-ph/0610277].
  %%CITATION = %doi:10.1016/j.physletb.2006.11.061;%%
  %118 citations counted in INSPIRE as of 10 Aug 2016
  
  \bibitem{Giudice:2011cg} 
  G.~F.~Giudice and A.~Strumia,
  %``Probing High-Scale and Split Supersymmetry with Higgs Mass Measurements,''
  Nucl.\ Phys.\ B {\bf 858}, 63 (2012)
 % doi:10.1016/j.nuclphysb.2012.01.001
  [arXiv:1108.6077 [hep-ph]].
  %%CITATION = doi:10.1016/j.nuclphysb.2012.01.001;%%
  %258 citations counted in INSPIRE as of 03 Sep 2018

  
  \bibitem{Hall:2011jd} 
  L.~J.~Hall and Y.~Nomura,
  %``Spread Supersymmetry,''
  JHEP {\bf 1201}, 082 (2012)
%  doi:10.1007/JHEP01(2012)082
  [arXiv:1111.4519 [hep-ph]].
  %%CITATION = doi:10.1007/JHEP01(2012)082;%%
  %148 citations counted in INSPIRE as of 03 Sep 2018

  
  
  \bibitem{pure2} 
  M.~Ibe and T.~T.~Yanagida,
  %``The Lightest Higgs Boson Mass in Pure Gravity Mediation Model,''
  Phys.\ Lett.\ B {\bf 709}, 374 (2012)
%  %doi:10.1016/j.physletb.2012.02.034
  [arXiv:1112.2462 [hep-ph]].
  %%CITATION = %doi:10.1016/j.physletb.2012.02.034;%%
    
    \bibitem{pure3} 
  N.~Arkani-Hamed, A.~Gupta, D.~E.~Kaplan, N.~Weiner and T.~Zorawski,
  %``Simply Unnatural Supersymmetry,''
  arXiv:1212.6971 [hep-ph].
  %%CITATION = ARXIV:1212.6971;%%
  
  
  
  
  
  
  \bibitem{polonyi1}
    G.~D.~Coughlan, W.~Fischler, E.~W.~Kolb, S.~Raby and G.~G.~Ross,
  %``Cosmological Problems for the Polonyi Potential,''
  Phys.\ Lett.\  {\bf 131B}, 59 (1983).
%  doi:10.1016/0370-2693(83)91091-2
  %%CITATION = doi:10.1016/0370-2693(83)91091-2;%%
  %659 citations counted in INSPIRE as of 03 Sep 2018
  
  
  
  
  \bibitem{ft_measure1}
  J.~R.~Ellis, K.~Enqvist, D.~V.~Nanopoulos and F.~Zwirner,
  %``Observables in Low-Energy Superstring Models,''
  Mod.\ Phys.\ Lett.\ A {\bf 1}, 57 (1986).
  %%CITATION = MPLAE,A1,57;%%
  \bibitem{ft_measure2}
   R.~Barbieri and G.~F.~Giudice,
  %``Upper Bounds on Supersymmetric Particle Masses,''
  Nucl.\ Phys.\ B {\bf 306}, 63 (1988).
  %%CITATION = NUPHA,B306,63;%%  
  
 
  
  %Higgs mass
\bibitem{Okada:1990vk}
  Y.~Okada, M.~Yamaguchi and T.~Yanagida,
  %``Upper bound of the lightest Higgs boson mass in the minimal supersymmetric standard model,''
  Prog.\ Theor.\ Phys.\  {\bf 85} 1 (1991).
  %doi:10.1143/ptp/85.1.1
  %%CITATION = doi:10.1143/ptp/85.1.1;%%
  %1329 citations counted in INSPIRE as of 01 May 2018

%\cite{Ellis:1990nz}
\bibitem{Ellis:1990nz}
  J.~R.~Ellis, G.~Ridolfi and F.~Zwirner,
  %``Radiative corrections to the masses of supersymmetric Higgs bosons,''
  Phys.\ Lett.\ B {\bf 257} 83 (1991).
  %doi:10.1016/0370-2693(91)90863-L
  %%CITATION = doi:10.1016/0370-2693(91)90863-L;%%
  %1481 citations counted in INSPIRE as of 01 May 2018

\bibitem{Haber:1990aw}
  H.~E.~Haber and R.~Hempfling,
  %``Can the mass of the lightest Higgs boson of the minimal supersymmetric model be larger than m(Z)?,''
  Phys.\ Rev.\ Lett.\  {\bf 66} 1815 (1991).
  %doi:10.1103/PhysRevLett.66.1815
  %%CITATION = doi:10.1103/PhysRevLett.66.1815;%%
  %1468 citations counted in INSPIRE as of 01 May 2018

\bibitem{Okada:1990gg}
  Y.~Okada, M.~Yamaguchi and T.~Yanagida,
  %``Renormalization group analysis on the Higgs mass in the softly broken supersymmetric standard model,''
  Phys.\ Lett.\ B {\bf 262} 54 (1991).
 % doi:10.1016/0370-2693(91)90642-4
  %%CITATION = doi:10.1016/0370-2693(91)90642-4;%%
  %600 citations counted in INSPIRE as of 01 May 2018

\bibitem{Ellis:1991zd}
  J.~R.~Ellis, G.~Ridolfi and F.~Zwirner,
  %``On radiative corrections to supersymmetric Higgs boson masses and their implications for LEP searches,''
  Phys.\ Lett.\ B {\bf 262} 477 (1991).
 % doi:10.1016/0370-2693(91)90626-2
  %%CITATION = doi:10.1016/0370-2693(91)90626-2;%%
  %799 citations counted in INSPIRE as of 01 May 2018



\bibitem{Fukuda:2015pra} 
  H.~Fukuda, H.~Murayama, T.~T.~Yanagida and N.~Yokozaki,
  %``Seminatural Gauge Mediation from Product Group Unification,''
  Phys.\ Rev.\ D {\bf 92}, no. 5, 055032 (2015)
%  doi:10.1103/PhysRevD.92.055032
  [arXiv:1508.00445 [hep-ph]].
  %%CITATION = doi:10.1103/PhysRevD.92.055032;%%
  %5 citations counted in INSPIRE as of 31 Aug 2018
  
  %\cite{Agashe:1999ct}
\bibitem{Agashe:1999ct} 
  K.~Agashe,
  %``Can multi - TeV (top and other) squarks be natural in gauge mediation?,''
  Phys.\ Rev.\ D {\bf 61}, 115006 (2000)
%  doi:10.1103/PhysRevD.61.115006
  [hep-ph/9910497].
  %%CITATION = doi:10.1103/PhysRevD.61.115006;%%
  %26 citations counted in INSPIRE as of 03 Sep 2018

  
  %\cite{Brummer:2012zc}
\bibitem{Brummer:2012zc} 
  F.~Brummer and W.~Buchmuller,
  %``The Fermi scale as a focus point of high-scale gauge mediation,''
  JHEP {\bf 1205}, 006 (2012)
%  doi:10.1007/JHEP05(2012)006
  [arXiv:1201.4338 [hep-ph]].
  %%CITATION = doi:10.1007/JHEP05(2012)006;%%
  %45 citations counted in INSPIRE as of 03 Sep 2018

\bibitem{Brummer:2013yya} 
  F.~Brümmer, M.~Ibe and T.~T.~Yanagida,
  %``Focus point gauge mediation in product group unification,''
  Phys.\ Lett.\ B {\bf 726}, 364 (2013)
 % doi:10.1016/j.physletb.2013.09.012
  [arXiv:1303.1622 [hep-ph]].
  %%CITATION = doi:10.1016/j.physletb.2013.09.012;%%
  %12 citations counted in INSPIRE as of 03 Sep 2018


 \bibitem{Yanagida:2013ah} 
  T.~T.~Yanagida and N.~Yokozaki,
  %``Focus Point in Gaugino Mediation ~ Reconsideration of the Fine-tuning Problem ~,''
  Phys.\ Lett.\ B {\bf 722}, 355 (2013)
  %doi:10.1016/j.physletb.2013.04.043
  [arXiv:1301.1137 [hep-ph]].
  %%CITATION = doi:10.1016/j.physletb.2013.04.043;%%
  %33 citations counted in INSPIRE as of 03 Sep 2018

\bibitem{Yanagida:2013uka} 
  T.~T.~Yanagida and N.~Yokozaki,
  %``Bino-Higgsino Mixed Dark Matter in a Focus Point Gaugino Mediation,''
  JHEP {\bf 1311}, 020 (2013)
  %doi:10.1007/JHEP11(2013)020
  [arXiv:1308.0536 [hep-ph]].
  %%CITATION = doi:10.1007/JHEP11(2013)020;%%
  %14 citations counted in INSPIRE as of 03 Sep 2018
  
  %\cite{Yanagida:2014cxa}
\bibitem{Yanagida:2014cxa} 
  T.~T.~Yanagida and N.~Yokozaki,
  %``Upper Bounds on Gluino, Squark and Higgisino Masses in the Focus Point Gaugino Mediation with a Mild Fine Tuning $\Delta \le 100$,''
  JHEP {\bf 1410}, 133 (2014)
 % doi:10.1007/JHEP10(2014)133
  [arXiv:1404.2025 [hep-ph]].
  %%CITATION = doi:10.1007/JHEP10(2014)133;%%
  %7 citations counted in INSPIRE as of 03 Sep 2018
  
  
  %\cite{Kane:1998im}
\bibitem{Kane:1998im} 
  G.~L.~Kane and S.~F.~King,
  %``Naturalness implications of LEP results,''
  Phys.\ Lett.\ B {\bf 451}, 113 (1999)
%  doi:10.1016/S0370-2693(99)00190-2
  [hep-ph/9810374].
  %%CITATION = doi:10.1016/S0370-2693(99)00190-2;%%
  %182 citations counted in INSPIRE as of 03 Sep 2018

\bibitem{BasteroGil:1999gu} 
  M.~Bastero-Gil, G.~L.~Kane and S.~F.~King,
  %``Fine tuning constraints on supergravity models,''
  Phys.\ Lett.\ B {\bf 474}, 103 (2000)
%  doi:10.1016/S0370-2693(00)00002-2
  [hep-ph/9910506].
  %%CITATION = doi:10.1016/S0370-2693(00)00002-2;%%
  %89 citations counted in INSPIRE as of 03 Sep 2018

%\cite{Abe:2007kf}
\bibitem{Abe:2007kf} 
  H.~Abe, T.~Kobayashi and Y.~Omura,
  %``Relaxed fine-tuning in models with non-universal gaugino masses,''
  Phys.\ Rev.\ D {\bf 76}, 015002 (2007)
%  doi:10.1103/PhysRevD.76.015002
  [hep-ph/0703044 [HEP-PH]].
  %%CITATION = doi:10.1103/PhysRevD.76.015002;%%
  %94 citations counted in INSPIRE as of 03 Sep 2018
  
  \bibitem{Martin:2007gf} 
  S.~P.~Martin,
  %``Compressed supersymmetry and natural neutralino dark matter from top squark-mediated annihilation to top quarks,''
  Phys.\ Rev.\ D {\bf 75}, 115005 (2007)
%  doi:10.1103/PhysRevD.75.115005
  [hep-ph/0703097 [HEP-PH]].
  %%CITATION = doi:10.1103/PhysRevD.75.115005;%%
  %120 citations counted in INSPIRE as of 03 Sep 2018

\bibitem{Baer:2007xd} 
  H.~Baer, A.~Mustafayev, H.~Summy and X.~Tata,
  %``Mixed Higgsino dark matter from a large SU(2) gaugino mass,''
  JHEP {\bf 0710}, 088 (2007)
%  doi:10.1088/1126-6708/2007/10/088
  [arXiv:0708.4003 [hep-ph]].
  %%CITATION = doi:10.1088/1126-6708/2007/10/088;%%
  %25 citations counted in INSPIRE as of 03 Sep 2018

\bibitem{Horton:2009ed} 
  D.~Horton and G.~G.~Ross,
  %``Naturalness and Focus Points with Non-Universal Gaugino Masses,''
  Nucl.\ Phys.\ B {\bf 830}, 221 (2010)
%  doi:10.1016/j.nuclphysb.2009.12.031
  [arXiv:0908.0857 [hep-ph]].
  %%CITATION = doi:10.1016/j.nuclphysb.2009.12.031;%%
  %88 citations counted in INSPIRE as of 03 Sep 2018
  
  \bibitem{Younkin:2012ui} 
  J.~E.~Younkin and S.~P.~Martin,
  %``Non-universal gaugino masses, the supersymmetric little hierarchy problem, and dark matter,''
  Phys.\ Rev.\ D {\bf 85}, 055028 (2012)
%  doi:10.1103/PhysRevD.85.055028
  [arXiv:1201.2989 [hep-ph]].
  %%CITATION = doi:10.1103/PhysRevD.85.055028;%%
  %54 citations counted in INSPIRE as of 03 Sep 2018
  
  \bibitem{Antusch:2012gv} 
  S.~Antusch, L.~Calibbi, V.~Maurer, M.~Monaco and M.~Spinrath,
  %``Naturalness of the Non-Universal MSSM in the Light of the Recent Higgs Results,''
  JHEP {\bf 1301}, 187 (2013)
%  doi:10.1007/JHEP01(2013)187
  [arXiv:1207.7236 [hep-ph]].
  %%CITATION = doi:10.1007/JHEP01(2013)187;%%
  %70 citations counted in INSPIRE as of 03 Sep 2018
  
  \bibitem{Gogoladze:2012yf} 
  I.~Gogoladze, F.~Nasir and Q.~Shafi,
  %``Non-Universal Gaugino Masses and Natural Supersymmetry,''
  Int.\ J.\ Mod.\ Phys.\ A {\bf 28}, 1350046 (2013)
%  doi:10.1142/S0217751X13500462
  [arXiv:1212.2593 [hep-ph]].
  %%CITATION = doi:10.1142/S0217751X13500462;%%

\bibitem{Spies:2013fba} 
  A.~Spies and G.~Anton,
  %``Confronting Recent Results from Selected Direct and Indirect Dark Matter Searches and the Higgs Boson with Supersymmetric Models with Non-universal Gaugino Masses,''
  JCAP {\bf 1306}, 022 (2013)
%  doi:10.1088/1475-7516/2013/06/022
  [arXiv:1306.1099 [hep-ph]].
  %%CITATION = doi:10.1088/1475-7516/2013/06/022;%%
  %4 citations counted in INSPIRE as of 03 Sep 2018

  
  \bibitem{Martin:2013aha} 
  S.~P.~Martin,
  %``Nonuniversal gaugino masses and seminatural supersymmetry in view of the Higgs boson discovery,''
  Phys.\ Rev.\ D {\bf 89}, no. 3, 035011 (2014)
%  doi:10.1103/PhysRevD.89.035011
  [arXiv:1312.0582 [hep-ph]].
  %%CITATION = doi:10.1103/PhysRevD.89.035011;%%
  %34 citations counted in INSPIRE as of 03 Sep 2018



\bibitem{Porod:2003um} 
  W.~Porod,
  %``SPheno, a program for calculating supersymmetric spectra, SUSY particle decays and SUSY particle production at e+ e- colliders,''
  Comput.\ Phys.\ Commun.\  {\bf 153}, 275 (2003)
%  doi:10.1016/S0010-4655(03)00222-4
  [hep-ph/0301101].
  %%CITATION = doi:10.1016/S0010-4655(03)00222-4;%%
  %821 citations counted in INSPIRE as of 03 Sep 2018

\bibitem{Porod:2011nf} 
  W.~Porod and F.~Staub,
  %``SPheno 3.1: Extensions including flavour, CP-phases and models beyond the MSSM,''
  Comput.\ Phys.\ Commun.\  {\bf 183}, 2458 (2012)
%  doi:10.1016/j.cpc.2012.05.021
  [arXiv:1104.1573 [hep-ph]].
  %%CITATION = doi:10.1016/j.cpc.2012.05.021;%%
  %451 citations counted in INSPIRE as of 03 Sep 2018
  
  
  \bibitem{stau_lhc}
The CMS Collaboration, CMS-PAS-EXO-16-036.

\bibitem{Baur:2015jsy} 
  J.~Baur, N.~Palanque-Delabrouille, C.~Yèche, C.~Magneville and M.~Viel,
  %``Lyman-alpha Forests cool Warm Dark Matter,''
  JCAP {\bf 1608}, no. 08, 012 (2016)
 % doi:10.1088/1475-7516/2016/08/012
  [arXiv:1512.01981 [astro-ph.CO]].
  %%CITATION = doi:10.1088/1475-7516/2016/08/012;%%
  %76 citations counted in INSPIRE as of 31 Aug 2018




\bibitem{Ellwanger:2008py} 
  U.~Ellwanger, C.-C.~Jean-Louis and A.~M.~Teixeira,
  %``Phenomenology of the General NMSSM with Gauge Mediated Supersymmetry Breaking,''
  JHEP {\bf 0805}, 044 (2008)
%  doi:10.1088/1126-6708/2008/05/044
  [arXiv:0803.2962 [hep-ph]].
  %%CITATION = doi:10.1088/1126-6708/2008/05/044;%%
  %44 citations counted in INSPIRE as of 30 Aug 2018

%\cite{Allanach:2015mwa}
\bibitem{Allanach:2015mwa} 
  B.~Allanach, M.~Badziak, C.~Hugonie and R.~Ziegler,
  %``Gauge Mediation in the NMSSM with a Light Singlet: Sparticles within the Reach of LHC Run II,''
  PoS PLANCK {\bf 2015}, 012 (2015)
  [arXiv:1510.03143 [hep-ph]].
  %%CITATION = ARXIV:1510.03143;%%
  %2 citations counted in INSPIRE as of 30 Aug 2018


%\cite{Aaboud:2017vwy}
\bibitem{Aaboud:2017vwy} 
  M.~Aaboud {\it et al.} [ATLAS Collaboration],
  %``Search for squarks and gluinos in final states with jets and missing transverse momentum using 36  fb$^{-1}$ of $\sqrt{s}=13$  TeV pp collision data with the ATLAS detector,''
  Phys.\ Rev.\ D {\bf 97}, no. 11, 112001 (2018)
%  doi:10.1103/PhysRevD.97.112001
  [arXiv:1712.02332 [hep-ex]].
  %%CITATION = doi:10.1103/PhysRevD.97.112001;%%
  %37 citations counted in INSPIRE as of 30 Aug 2018
  


\bibitem{Yanagida:1994vq} 
  T.~Yanagida,
  %``Naturally light Higgs doublets in the supersymmetric grand unified theories with dynamical symmetry breaking,''
  Phys.\ Lett.\ B {\bf 344}, 211 (1995)
  %doi:10.1016/0370-2693(94)01500-C
  [hep-ph/9409329].
  %%CITATION = doi:10.1016/0370-2693(94)01500-C;%%
  %93 citations counted in INSPIRE as of 03 Sep 2018

\bibitem{Izawa:1997he} 
  K.~I.~Izawa and T.~Yanagida,
  %``R invariant natural unification,''
  Prog.\ Theor.\ Phys.\  {\bf 97}, 913 (1997)
%  doi:10.1143/PTP.97.913
  [hep-ph/9703350].
  %%CITATION = doi:10.1143/PTP.97.913;%%
  %54 citations counted in INSPIRE as of 03 Sep 2018
  
  
  

  

%\cite{Aaboud:2017hrg}
\bibitem{Aaboud:2017hrg} 
  M.~Aaboud {\it et al.} [ATLAS Collaboration],
  %``Search for supersymmetry in final states with missing transverse momentum and multiple $b$-jets in proton-proton collisions at $ \sqrt{s}=13 $ TeV with the ATLAS detector,''
  JHEP {\bf 1806}, 107 (2018)
%  doi:10.1007/JHEP06(2018)107
  [arXiv:1711.01901 [hep-ex]].
  %%CITATION = doi:10.1007/JHEP06(2018)107;%%
  %12 citations counted in INSPIRE as of 30 Aug 2018


\end{thebibliography}
\end{document}